\documentclass[journal]{IEEEtran}
\usepackage{cite}
\usepackage{mciteplus}
\usepackage{graphicx}
\usepackage{bm}
\usepackage[dvipsnames]{xcolor}
\usepackage{subfigure}

\usepackage{caption}
\usepackage{breqn}
\usepackage{xspace} 
\usepackage{algorithm,algorithmic}

\usepackage{cellspace}
\usepackage{tabularx,booktabs,caption,ragged2e}
\usepackage{xfrac}
\usepackage{amsmath}
\usepackage{amssymb}
\usepackage{cancel}
\usepackage[normalem]{ulem}
\usepackage{subcaption}
\usepackage{float}
\usepackage{verbatim}
\usepackage{acro}
\DeclareAcronym{RSMA}{short = RSMA, long = rate splitting multiple access}

\DeclareAcronym{SIC}{short = SIC, long = successive interference cancellation}

\DeclareAcronym{BER}{short = BER, long = bit error rate}
\usepackage[hidelinks,colorlinks=true,linkcolor=blue,citecolor=blue,urlcolor=black]{hyperref}
\begin{document}

\title{Waveform Coexistence-Driven RSMA: \\ A Pioneering Strategy for Future 6G Networks}

\author{Kenza Abela, Shaima Abidrabbu, Ayoub Ammar Boudjelal, and H\"{u}seyin Arslan, {\it Fellow, IEEE}
\thanks{The authors are with Department of Electrical and Electronics Engineering, Istanbul Medipol University, 34810 Istanbul, Türkiye. (e-mail: abela.kenza@std.medipol.edu.tr, shaima.abidrabbu@std.medipol.edu.tr, 
ayoub.ammar@std.medipol.edu.tr, huseyinarslan@medipol.edu.tr).}
\thanks{S. Abidrabbu, is with IPR and License Agreements Department, Vestel Electronics,  45030 Manisa, Turkey. (e-mail: shaima.abidrabbu@vestel.com.tr).}\protect}

\maketitle
\begin{abstract}
Two critical approaches have emerged in the literature for the successful realization of 6G wireless networks: the coexistence of multiple waveforms and the adoption of non-orthogonal multiple access. These strategies hold transformative potential for addressing the limitations of current systems and enabling the robust and scalable design of next-generation wireless networks. This paper presents a novel rate splitting multiple access (RSMA) framework that leverages the coexistence of affine frequency division multiplexing (AFDM) and orthogonal frequency division multiplexing (OFDM). By transmitting common data via AFDM at higher power in the affine domain and private data via OFDM at lower power in the frequency domain, the proposed framework eliminates the reliance on successive interference cancellation (SIC), significantly simplifying receiver design. Furthermore, two data mapping approaches are proposed: a clean pilot method, where pilots are allocated without any data overlapping, ensuring clear separation, and an embedded pilot method, where pilots overlap with data for more efficient resource utilization. Channel estimation is then performed for different channel types. Simulation results demonstrate the robustness and efficiency of the proposed approach, achieving superior performance in efficiency, reliability, and adaptability under diverse channel conditions. This framework transforms non-orthogonal multi-access design, paving the way for scalable and efficient solutions in 6G networks.
\end{abstract}


%\IEEEpeerreviewmaketitl

\section{INTRODUCTION}
\IEEEPARstart{N}EXT-GENERATION multiple access (NGMA) techniques have emerged as critical enablers to address the complex challenges posed by next-generation networks \cite{liu2022developing}. Unlike traditional approaches such as orthogonal multiple access (OMA), which allocate resources rigidly, or non-orthogonal multiple access (NOMA)\cite{8357810}, which depends heavily on SIC, NGMA introduces a paradigm shift by prioritizing flexibility and adaptability. These novel strategies aim to maximize spectral efficiency (SE), ensure fairness, and deliver robust performance in highly dynamic and diverse environments. Among NGMA techniques, RSMA stands out as a promising candidate for revolutionizing resource and interference management in multi-user systems. RSMA achieves this by dividing user data into a common component, shared by multiple users, and private components, specific to individual users \cite{mao2022rate}. This approach transforms interference from a limitation into a resource, leveraging it to optimize performance. By employing linear precoding, RSMA ensures superior SE, fairness, outperforming established schemes such as NOMA and Space division multiple access (SDMA) \cite{mao2018rate}. 

In parallel with advancements in multiple access techniques, 6G networks are expected to adopt a hybrid waveform paradigm to address their diverse and stringent requirements. Unlike earlier generations that relied on a single waveform, 6G envisions the coexistence of complementary waveforms, each tailored for specific use cases \cite{hammoodi2019green}. For example, OFDM excels in delivering high data rates and straightforward equalization in frequency-selective channels \cite{468055}, while AFDM offers superior resilience against Doppler effects and time-frequency dispersion. By integrating such waveforms, 6G can achieve robust performance across high-mobility scenarios, dense urban environments, and integrated sensing and communication (ISAC) use cases. This coexistence enhances flexibility in resource allocation, boosts spectral efficiency, and improves resilience to channel impairments, making it a cornerstone of 6G systems \cite{demir2024waveform}. However, existing frameworks fail to synergize these waveforms with NGMA techniques, creating a critical research gap. Addressing this, our work introduces a novel RSMA architecture that leverages the coexistence of these complementary waveforms to redefine multi-access paradigms for 6G.

\subsection{Related Literature}
Several studies on RSMA have extensively explored resource allocation, precoder optimization, and multi-carrier waveform design. However, many of these works, such as \cite{9201435,9663192,9403732,9541482,8926413,9705574}, focus primarily on theoretical aspects, often overlooking practical waveform impairments and their impact on system performance. A key limitation is the reliance on OFDM, a waveform highly susceptible to intercarrier interference in high-mobility scenarios or under Doppler shifts. Although RSMA offers potential improvements in spectral efficiency and robustness, the existing literature lacks exploration of innovative waveform designs to mitigate these vulnerabilities.

Recent studies have begun addressing these gaps by integrating advanced techniques into RSMA. For example, multi-numerology OFDM-RSMA schemes have been proposed to mitigate intercarrier interference and improve sum-rate performance through optimized power and subcarrier allocation \cite{10236464,10516586}. These approaches outperform traditional methods like OFDMA and OFDM-NOMA under diverse channel conditions, underscoring RSMA’s adaptability to waveform~impairments.

Furthermore, RSMA has been extended to high-mobility environments using orthogonal time frequency space (OTFS) modulation. For uplink scenarios, \cite{huai2024cross} proposes an OTFS-RSMA cross-domain scheme that employs heterogeneous user grouping to enhance outage performance and user fairness. Similarly, \cite{10804646} develops an OTFS-RSMA transmission scheme with a power allocation strategy that optimizes the average sum rate using a successive convex approximation algorithm, achieving superior performance compared to OTFS-NOMA and OFDM under high mobility. In the downlink, \cite{huai2025downlink} introduces a cross-domain OTFS-RSMA scheme using heterogeneous user grouping and multicast technology, achieving substantial user sum-rate gains over OTFS-NOMA.
RSMA's applicability has also been explored in non-terrestrial networks. For example, \cite{10741218} tackles the precoder design bottleneck by employing a deep learning-based method to predict precoder designs directly from historical CSI, reducing computational complexity and feedback requirements. Additionally, practical implementations of RSMA have been demonstrated. \cite{10471302} presents an RSMA prototype using software-defined radios, validating RSMA’s superior sum throughput and fairness over SDMA and NOMA in real-world scenarios.

Addressing the challenges of SIC, studies have explored RSMA designs that eliminate its reliance. For instance, the Quadrature-RSMA (Q-RSMA) framework introduced in \cite{9831449} leverages quadrature carriers to avoid SIC for private streams while enabling interference-free detection of common streams, significantly improving spectral efficiency and decoding latency. Similarly, the authors of \cite{10485496} investigate SIC-free RSMA designs for finite-alphabet constellations, demonstrating RSMA’s resilience against SDMA even without SIC. In another empirical study \cite{10694543}, the authors evaluate RSMA implementations using software-defined radios, exploring trade-offs between single and multiple common message structures and confirming the feasibility of robust system designs.

\subsection{Motivation and Contributions}
Despite advancements in RSMA research, most studies focus on a single waveform, primarily OFDM, which restricts the exploration of alternative methods that could address these issues. To overcome this limitation, this work proposes a novel RSMA framework based on the coexistence of two complementary waveforms: AFDM and OFDM. By eliminating SIC and utilizing the distinct domains of these waveforms for data separation, the proposed framework enhances spectral efficiency, reduces complexity, and offers a flexible, robust solution for next-generation wireless systems. The following are the main contributions of this work:

\begin{itemize}
\item A new RSMA framework that leverages the coexistence of two complementary waveforms (AFDM and OFDM) is proposed. AFDM is employed for high-power common data transmission due to its robustness against time-frequency dispersive channels, while OFDM is utilized for low-power private data transmission, benefiting from its SE and compatibility with multi-antenna systems.
\item The proposed framework introduces domain-based data stream separation, with AFDM operating in the affine domain for common data and OFDM in the frequency domain for private data. This separation enables the receiver to independently recognize and decode each stream, eliminating the need for SIC. 

\item To optimize channel estimation in the framework, two complementary strategies are proposed. The first employs dedicated clean pilot signals in the affine domain to achieve precise and interference-free channel estimation. The second incorporates pilot signals alongside common data in the affine domain, improving resource efficiency and boosting effective data rates. 

\item Simulation results validate the analytical findings and benchmark the proposed framework against state-of-the-art schemes. Key performance metrics, including \ac{BER} and SE, demonstrate the framework's superiority, showcasing its robustness, adaptability, and practicality for next-generation wireless networks.
\end{itemize}

\subsection{Organization}
The rest of the paper is organized as follows. Section II presents the system model along with preliminary concepts underlying the proposed approach. Section III provides a detailed analysis of the proposed method. Channel estimation techniques are explored in Section IV. Section V presents Analytical and simulation results. Finally, conclusions along with potential directions for future research are outlined in Section VI.

\textit{Notations:} Bold lowercase ($\mathbf{a}$) and uppercase ($\mathbf{A}$) letters denote vectors and matrices, respectively, while non-bold letters ($a$, $A$) represent scalars. The complex and real fields are indicated by $\mathbb{C}$ and $\mathbb{R}$, with $\mathbb{C}^{M \times N}$ defining the space of complex-valued matrices. $[\cdot]_{b}$ denoting the modulo operation. $\mathcal{F}\{\cdot\}$ and $\mathcal{F}^{-1}\{\cdot\}$ indicate transforms from affine to frequency domain and from frequency to affine domain, respectively. 

\section{SYSTEM MODEL AND PRELIMINARIES OF THE PROPOSED APPROACH}
In this section, the signaling of the conventional RSMA along with the conventional OFDM and AFDM, are presented. 
\begin{figure*}
    \centering
    \includegraphics[width=0.85\linewidth]{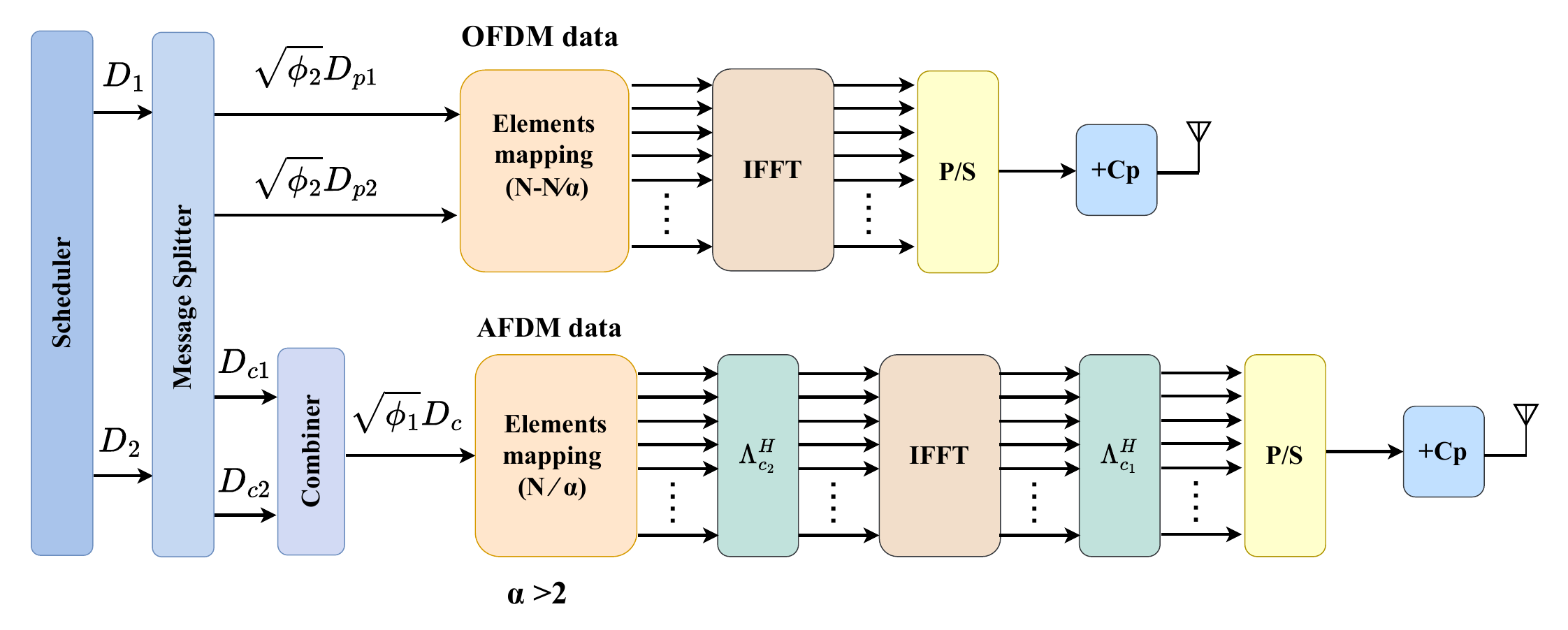}
    \caption{The proposed RSMA system model.}
    \label{fig1}
\end{figure*} 
\subsection{RSMA Signaling}
Consider a downlink RSMA network with one source equipped with $N_t$ transmitting antennas, serving $K$ users, each with a single receiving antenna. The source has a set of messages to be transmitted to the users, denoted as \{$W_1$,\ldots, $W_k$, \ldots, $W_K$\}. The message $W_k$ intended for the $k$-th user is split into two parts: common and private denoted as $W_{\mathrm{c}, k}$ and $W_{\mathrm{p},k}$, respectively, where each stream caters to data each with its own set of priorities. The common components from all users $\left\{ {W_{\mathrm{c}, 1}, \ldots, W_{\mathrm{c}, K}} \right\}$ are combined into one common message $W_{c}$, which is encoded into a common stream $s_{\mathrm{c}}$. This common stream is decoded by all the users in RSMA network. In contrast, the private components, ${W_{\mathrm{p}, 1}, \ldots, W_{\mathrm{p}, K}}$, are individually encoded into separate private streams ${s_1, \ldots, s_K}$ which are decoded by the intended users only. The data streams are denoted together by $\mathbf{s} = \left[s_{\mathrm{c}}, s_1, \ldots, s_K\right]^T \in \mathbb{C}^{K+1}$.  Consequently, the signal transmitted to the $k$-the user can be expressed as follows
\begin{equation}
    \textbf{s}=  \sqrt{p_{c}P_{t}} 
 {\textbf{p}_{c}} s_{\mathrm{c}} +  \sum_{k=1}^{K} \sqrt{p_{k}P_{t}} {\textbf{p}_{k}} s_{k}, \quad \forall k \in {K}, 
\end{equation}
where  ${\textbf{p}_{c}}, {\textbf{p}_{k}} \in \mathbb{C}^{N_t \times 1}$ are the precoders for the common stream and the $k$-th private stream, respectively. Moreover,  $p_{c}$ and  $p_{k}$ represent the power allocation coefficients for both streams and $P_{t}$ represents the total transmission power.
The received signal at the $k$-th user is given as 
\begin{equation}
  {y}_{k}=\mathbf{h}_{k}^{H}  \textbf{s}+{n}_{k}, 
\end{equation} where $\mathbf{h}_{k} \in \mathbb{C}^{N_{t} \times 1}$ denotes the channel gain between the $k$-th user and source which is fully known at the source, and ${n}_{k}$ represents the $k$-th user additive white Gaussian noise (AWGN) with zero mean and variance $\sigma_{k}^{2}$, expressed as ${n}_{k} \sim \mathcal{CN}\left(0,\sigma_{k}^{2}\right)$.
Assuming that $\mathbb{E}\left\{\textbf{ss}^H\right\}=\mathbf{I}$, the transmitted power at the source is constrained by $\operatorname{tr}\left(\mathbf{P} \mathbf{P}^H\right) \leq P_{\mathrm{t}}$  such that, $\mathbf{P} = \left[{\textbf{p}_{c}}, {\textbf{p}_{1}}, \ldots, {\textbf{p}_{K}}\right]$. Therefore, the power allocation coefficients are subject to the constraint $p_{c} + \sum_{k=1}^{K} p_{k} \leq P_{\mathrm{t}}$.

\subsection{OFDM Signaling} 
OFDM is designed to transmit high-data-rate streams by dividing them into multiple parallel low-rate streams, each transmitted over orthogonal subcarrier. The input data symbols, denoted as \( X^{\text{OFDM}} = \{x_0, x_1, \ldots, x_{N-1}\} \), are modulated onto \( N \) orthogonal subcarriers and transmitted in parallel.
The OFDM time-domain signal can be expressed as
\begin{align}
    s^{\text{OFDM}}(n) = & \frac{1}{\sqrt{N}} \sum_{k=0}^{N-1} X^{\text{OFDM}}(k) e^{j 2 \pi \frac{k n}{N}}, \notag \\
    & \quad n = 0, 1, \ldots, N-1.
\end{align}
 To combat inter-symbol interference (ISI) caused by multipath fading, a cyclic prefix (CP) is added before transmission.

\subsection{AFDM Signaling}
AFDM is an advanced multicarrier waveform that builds on traditional techniques like OFDM but introduces unique chirp-based modulation. In AFDM, the signal is constructed as a combination of multiple chirp signals using the Inverse Discrete Affine Fourier Transform (IDAFT). 
The AFDM signal is expressed as~\cite{10087310}
\begin{equation}
s^{\text{AFDM}}(n) = \frac{1}{\sqrt{N}} \sum_{i=0}^{N-1} X^{\text{AFDM}}(i) e^{j 2 \pi \left(c_1 n^2  + c_2 i^2 + \frac{n i}{N} \right)},
\end{equation}
where \( X^{\text{AFDM}} (i) \) represents the data symbol modulated onto the $i^{th}$ twisted chirp, \( c_1 \) controls the chirp rate in the time domain, \( c_2 \) adjusts the chirp rate across frequency domain.
The affine transformation parameters \( c_1 \) and \( c_2 \) spread symbols across Time-Frequency (TF) resources, enhancing resilience against delay and Doppler shifts introduced by the channel.

\begin{figure*}[t]
    \centering
    \includegraphics[width=1\textwidth]{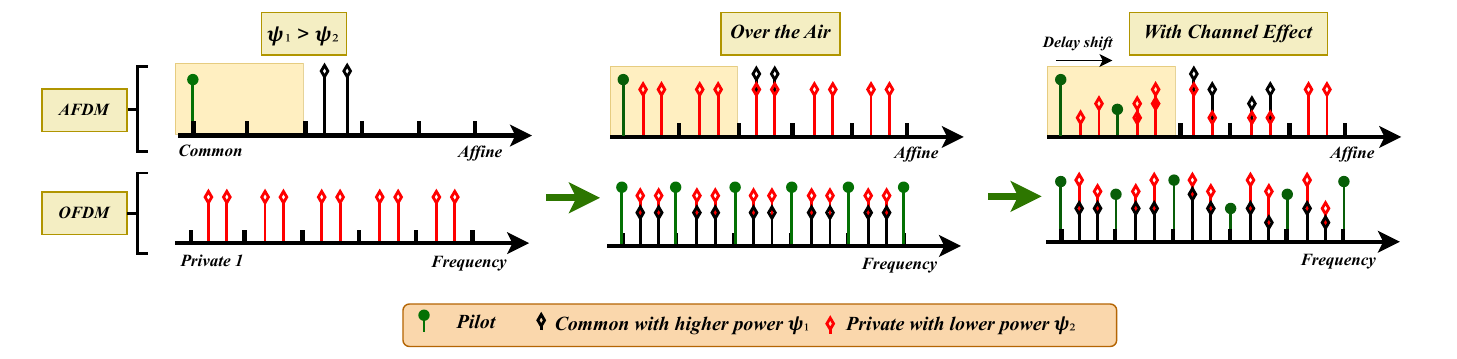}
    \caption{First Approach (Clean pilot approach).}
    \label{fig2}
\end{figure*} 
\section{Proposed AFDM-OFDM-RSMA}
The main approach of this work is to generate two signals from two different waveforms, where each signal remains localized in its respective domain while being spread in the other signal's domain. This approach enables dual separation of the signals as a form of waveform-domain separation without requiring  SIC. In this framework, AFDM is utilized for its robustness against doubly-selective channels, while OFDM is chosen for its ability to be spread in the affine domain under specific conditions, as described in this section. The common data stream is transmitted using AFDM at higher power in the affine domain, whereas the private data stream is transmitted using OFDM at lower power in the frequency domain.

To illustrate the proposed strategy, we present a new system model in Fig. \ref{fig1}, where a base station (BS) employs RSMA to serve multiple users. For simplicity, the model initially considers only two users, user 1 and user 2, but the technique can be scaled to accommodate additional users. The process begins with a scheduler managing both users' data, assigning user 1's data (\(D_1\)) and user 2's data (\(D_2\)) within the system. A message splitter then divides the data into two components: the common data, shared among all users and handled via AFDM, and the private data, unique to each user and processed using OFDM. This approach allocates more data to the private portion since common data spreads across the frequency domain based on the affine translation factor (\(c_1\)), leading to a lower power density. To counteract this, the common data are transmitted at higher power, while the private data, being localized in the frequency domain, are sent at lower power. This power-level distinction, combined with domain-specific processing, facilitates robust decoding and enhances receiver performance.

\subsection{AFDM to OFDM (Frequency domain)}
\label{AFDM-OFDM}
To analyze the AFDM signal carrying the common data, its frequency domain representation is studied. The discrete-time AFDM signal after taking its Fast Fourier Transform (FFT) can be given in the frequency domain as
\begin{align}
    D_c^{\text{freq}}(m)& =  \sum_{n=0}^{N-1} s^{\text{AFDM}}(n) e^{-j\frac{2\pi nm}{N}}\\
    &
    =\frac{1}{\sqrt{N}} \sum_{n=0}^{N-1} \sum_{i=0}^{N-1} D_c(i) \notag \\
    & \cdot e^{j2\pi \left(c_1 n^2 + \frac{ni}{N} + c_2 i^2 \right)} e^{-j\frac{2\pi nm}{N}}.
\end{align}
After simplification, \(D_c^{\text{freq}}(m)\) can be written as
\begin{align} \label{11}
    D_c^{\text{freq}}(m) = & \frac{\sqrt{\phi_1}}{\sqrt{N}} \sum_{i=0}^{N-1} D_c(i) e^{j2\pi c_2 i^2} \notag \\
    & \cdot \sum_{n=0}^{N-1} e^{j2\pi \left(c_1 n^2 + \frac{n(i-m)}{N}\right)}.
\end{align}
By setting \(c_1 = \frac{c_1'}{2N}\) such that \(c_1'\) is an integer power of 2 in the second term of (\ref{11}), we can rewrite it as follows
\begin{equation}\label{12}
 \sum_{n=0}^{N-1} e^{j\pi \frac{c_1' n^2}{N}} e^{j\frac{2\pi ni}{N}} e^{-j\frac{2\pi nm}{N}}.   
\end{equation}
Therefore, (\eqref{12}) is giving exactly the Fourier transform of a chirp signal with slope \(c_1'\). Given \(N = c_1' M\) and \(M\) is a power of 2, a \(c_1'\)-repeated chirp is expressed as follows
\begin{equation}
 s^{\text{chirp}}(p) = e^{j\pi \frac{p^2}{M}}, \quad p = 0, \dots, M-1,   
\end{equation}
by setting \(n = p + cM\), where \(c = 0, \dots, c_1'-1\), the Fourier transform of \(c_1'\)-repeated chirp is represented as
\begin{equation}\label{14}
 \sum_{c=0}^{c_1'-1} \sum_{p=0}^{M-1} s^{\text{chirp}}(p) e^{j\frac{2\pi (p + cM)i}{N}} e^{-j\frac{2\pi m(p + cM)}{N}},   
\end{equation}
where \(i\) is the frequency shift and after doing some simplification, (\ref{14}) simplifies to
\begin{equation}
 \sum_{c=0}^{c_1'-1} e^{j\frac{2\pi (i - m' c_1')c}{c_1'}} \sum_{p=0}^{M-1} e^{j\pi \frac{p^2}{M}} e^{j\frac{2\pi ip}{c_1' M}} e^{-j\frac{2\pi m'p}{M}},   
\end{equation}
where \(m' c_1' = m\) and \(m' = 0, \dots, M-1\).

As can be observed that the first term is non-zero only if \([i - m' c_1']_{c_1'} = \alpha\), which implies \([i]_{c_1'} = \alpha\), where \(\alpha = 0,\dots, \alpha\).

\begin{equation}
   \sum_{c=0}^{c_1'-1} e^{j\frac{2\pi (i - m' c_1')c}{c_1'}} =
\begin{cases} 
c_1', & \text{if } [m]_{c_1'} = \alpha \text{ and } [i]_{c_1'} = \alpha, \\
0, & \text{otherwise.}
\end{cases} 
\end{equation}

\begin{figure*}[ht]
    \centering
    \includegraphics[width=1\textwidth]{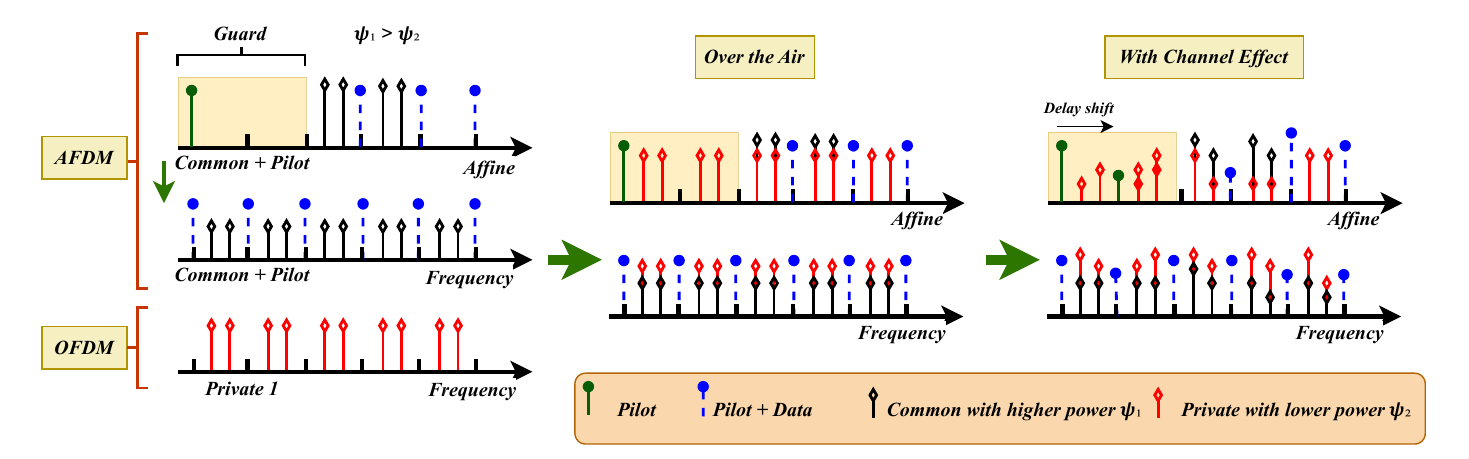}
    \caption{Second Approach (Data and pilot approach).}
    \label{fig3}
\end{figure*}

Similarly, as can be observed that the second term is the Fourier transform of a single chirp with rate \(1/M\) shifted by \(e^{j\frac{2\pi ip}{c_1' M}}\). As a result, the final expression of the AFDM signal in the frequency domain is presented as
\begin{equation}
D_c^{\text{freq}}(m) = \frac{ c_1'}{\sqrt{N}} D_c(i) e^{j2\pi c_2 i^2} e^{-j\pi \frac{m^2}{M}} \phi^i(m).   
\end{equation}
where
\begin{equation}
\phi^i(m) = \sum_{p=0}^{M-1} e^{-j2\pi \frac{(p - m')^2}{M}} e^{j\frac{2\pi ip}{c_1' M}}.    
\end{equation}
As depicted in Fig.~\ref{fig2} and Fig.~\ref{fig3}, the AFDM is spread in the frequency domain such that each \([i]_{c_1'} = \alpha\) contributes to each \([m]_{c_1'} = \alpha\) subcarrier.

\subsection{OFDM to AFDM (Affine domain)}
To represent the aggregated data in the affine domain, the OFDM signal is analyzed using the discrete affine Fourier transform (DAFT). Consequently, the OFDM private data is mapped into the Fresnel domain as follows
\begin{align}\label{19}
    D_{p1}^{\text{affine}}(i) = & \frac{1}{\sqrt{N}} \sum_{m=0}^{N-1} D_{p1}(m) e^{-j2\pi c_2 i^2} \notag \\
    & \cdot \sum_{n=0}^{N-1} e^{-j2\pi \left(\frac{c_1 N n^2 + n(i-m)}{N} \right)}.
\end{align}
Given the same conditions on \(n, m, i\), and \(c_1\) as mentioned earlier, (\ref{19}) can be written as
\begin{equation}
D_{p1}^{\text{affine},m}(i) =
\begin{cases}
    \frac{ c_1'}{\sqrt{N}} D_{p1}(m) 
    e^{-j2\pi c_2 i^2} e^{+j\pi \frac{m^2}{M}} \left(\phi^i(m)\right)^*, & \\[10pt]
    \quad [m]_{c_1'} = \alpha \text{ and } [i]_{c_1'} = \alpha, & \\[10pt]
    0, & \hspace{-3cm}\text{otherwise}.
\end{cases}
\end{equation}
where \(\left(\phi^i(m)\right)^*\) is the conjugate of \(\phi^i(m)\). Therefore, OFDM is spread in the affine domain such that each \([m]_{c_1'} = \alpha\) subcarrier contributes to each \([i]_{c_1'} = \alpha\) AFDM element, as depicted in Fig. \ref{fig2} and \ref{fig3}.

\subsection{Proposed approach 1 (Clean Pilot)} 
Leveraging affine and frequency domain properties of dual sparsity,  both AFDM and OFDM can be used to carry data with different powers and different number of data. The common data $D_c^{\text{affine}}$ are mapped in the affine domain at indices \(([i]_{c_{1}^{'}} \neq 0,~ G<i<N-G)\) where $G$ is a guard left following the channel state \cite{9473655} as follows 
\begin{equation}
   S_c^{\text{affine}}(i) =
\begin{cases} 
\sqrt{\phi_1} D_c^{\text{affine}}(i), & \text{if } i>G~ \text{and}[i]_{c_1'} \neq 0, \\
0, & \text{otherwise},
\end{cases} 
\label{equ:affine_com}
\end{equation}
where $\sqrt{\phi_1}$ is the power allocated to the affine domain data. An affine domain pilot $P^{\text{affine}}$  is inserted as follows 
\begin{equation}
   S_{pilot}^{\text{affine}}(i) =
\begin{cases} 
\sqrt{\phi} P^{\text{affine}}(i), & i=0, \\
0, & \text{otherwise},
\end{cases} 
\label{equ:affine_pilot}
\end{equation}
where $ \sqrt{\phi}$ is the power assigned to the pilot.
Since the affine domain elements are spreading in the frequency domain, the common data $S_c^{\text{affine}}$ will spread in the frequency domain following the process in section \ref{AFDM-OFDM} thus contributing to the subcarriers indices $[m]_{c_1^{\prime}} \neq 0$ while the affine domain pilot will spread in the subcarriers following $[m]_{c_1^{\prime}} = 0$ which can be used also for frequency domain channel estimation and equalization. The private data can be allocated to the frequency indices $[m]_{c_1^{\prime}} \neq 0$ to keep the pilot indices clean as follows 
\begin{equation}
   S_{p1}^{\text{freq}}(m) =
\begin{cases} 
\sqrt{\phi_2} D_{p1}^{\text{freq}}(m), & [m]_{c_1^{\prime}} \neq 0, \\
0, &\text{otherwise},
\end{cases} 
\label{equ:frequency_pri}
\end{equation}
where $\sqrt{\phi_2}$ is the power assigned to the private data. Finally, the combined signal can be expressed in frequency domain as follows

\begin{equation}
   S_{\text{comb1}}^{\text{freq}}(m) = \\
\begin{cases} 
\frac{ \mathcal{F}\{S_c^{\text{affine}}\}}{\sqrt{N}} (m)+\sqrt{\phi_2}D_{p1}^{\text{freq}}(m), & [m]_{c_1'} \neq 0, \\
\frac{ \mathcal{F}\{S_{pilot}^{\text{affine}}\}}{\sqrt{N}} (m), & [m]_{c_1'}=0,
\end{cases} 
\label{equ:approach1_comb}
\end{equation}
where $\mathcal{F}\{\cdot\} $ is the transform mapping the affine domain data to the frequency domain. Note that by using less common data than the private data with power $\sqrt{\phi_1}>\sqrt{\phi_2}$, the separability of both data is preserved. The combined signal is transformed to time domain signal $s_{\text{comb1}}^{\text{time}}$ using Inverse Fast Fourier Transform (IFFT) as follows 
\begin{equation}
    s_{\text{comb1}}^{\text{time}}(n)=\frac{1}{\sqrt{N}} \sum_{k=0}^{N-1} S_{\text{comb1}}^{\text{freq}}(m)  e^{j 2 \pi \frac{m n}{N}}.
\end{equation}

\begin{figure*}
    \centering
    \includegraphics[width=1\linewidth]{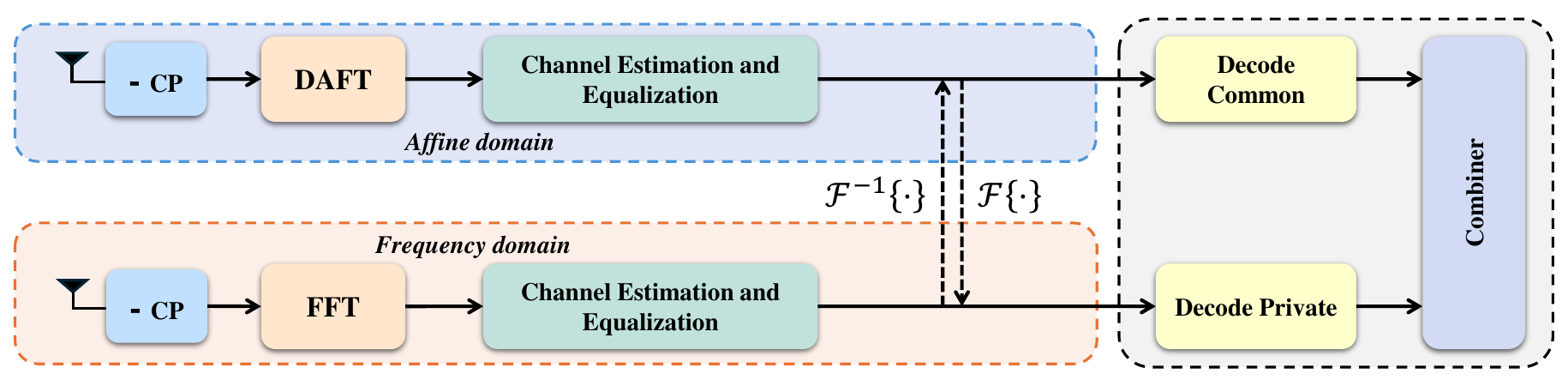}
    \caption{The proposed receiver design.}
    \label{fig4}
\end{figure*}

\subsection{Proposed approach 2 (Pilot and Data)}  
Unlike the proposed Approach 1, this approach allocates extra affine data $S_{ce}^{\text{affine}}$ in the indices \([i]_{c_{1}^{'}}=0 ,~ G<i<N-G)\) as depicted in Fig. \ref{fig3}, where these extra data are allocated directly with no extra power since they will not overlap with the private data in any domain. However, this allocation will lead to an overlap between the pilot and these data in the frequency domain. Therefore, the combined signal can be expressed in frequency domain as follows 
\begin{equation}
   S_{\text{comb2}}^{\text{freq}}(m) = \\
\begin{cases} 
\frac{ \mathcal{F}\{S_c^{\text{affine}}\}}{\sqrt{N}} (m)+\sqrt{\phi_2}D_{p1}^{\text{freq}}(m), & [m]_{c_1'} \neq 0, \\
\frac{ \mathcal{F}\{S_{pilot}^{\text{affine}}\}}{\sqrt{N}} (m)+ \frac{ \mathcal{F}\{S_{ce}^{\text{affine}}\}}{\sqrt{N}} (m), & [m]_{c_1'}=0,
\end{cases} 
\label{equ:approach2_comb}
\end{equation}
Note that the pilot in \eqref{equ:approach2_comb} is no longer clean, indicating the need for a different channel estimation method for this approach. The combined signal is transformed to time domain signal $s_{\text{comb2}}^{\text{time}}$ using IFFT as follows 
\begin{equation}
    s_{\text{comb2}}^{\text{time}}(n)=\frac{1}{\sqrt{N}} \sum_{k=0}^{N-1} S_{\text{comb2}}^{\text{freq}}(m)  e^{j 2 \pi \frac{m n}{N}}.
\end{equation}

\section{Channel Estimation}
 Consider an $R$-tap time-domain channel matrix\footnote{The delay and Doppler shifts are assumed to be integers.}. Specifically, $\mathbf{H}$ is defined as
\begin{equation}
    \mathbf{H} = \sum_{r=0}^{R-1} h_{r} \boldsymbol{\Pi}^{l_{r}}_{MN} \boldsymbol{\Delta}^{k_{r}}_{MN} + \mathbf{w},
    \label{equ:Channel_gen}
\end{equation}
where $l_r$, $k_r$, and $h_r$ denote the delay, Doppler shifts, and channel coefficients of the $r$-th path, respectively. $\mathbf{w}$ denotes the AWGN vector with variance $\sigma^2$. The matrix $\boldsymbol{\Pi}_{MN}^{l_r}$ is a permutation matrix with elements $\Pi_{MN}^{l_r}(k, l) = \delta([l-k]_{MN} - l_r)$, and $\boldsymbol{\Delta}_{MN}^{k_{r}}$ is an $M N \times M N$ diagonal matrix defined as
\begin{equation}
    \boldsymbol{\Delta}_{MN}^{k_{r}} = \operatorname{diag}\left[z_r^{0}, z_r^{1}, \ldots, z_r^{MN-1}\right],
\end{equation}
where $z_r = e^{\frac{j 2 \pi k_{r}}{M N}}$. The described channel can also be described in time by 
\begin{equation}
h(n) = \sum_{r=0}^{R-1} h_r e^{-\frac{j2\pi k_r (n-l_r)}{N}} \delta(n-l_r).  
\end{equation}

Therefore, the existence and the absence of Doppler require different processing, which will be discussed in both methods. In this section, two channel cases are considered; one tap delay channel and doubly-dispersive channel. 

\subsection{One Tap Delay Channel}
In the case of only delay in the environment, the channel equation can be represented by setting all $k_r=0$, as follows
\begin{equation}
 h(n) = \sum_{r=0}^{R-1} h_r \delta(n-l_r),   
\end{equation}
where each $l_r$ introduces a phase in the frequency domain and a shift with $c_1' l_r$ in the affine domain. After passing through the channel, the frequency domain of the received signal in approach 1  $Y1(m)$ can be represented as follows
\begin{equation} \label{24}
  Y(m) = S_{\text{comb1}}^{\text{freq}}(m) \cdot H(m),  
\end{equation}
where $H(m)$ is the channel frequency response. By substituting (\eqref{equ:approach1_comb}) and $H(m)$, the received pilot can be written as follows 

\begin{equation}
   Y1_{pilot}^{\text{freq}}(m) =\sum_{r=0}^{R-1} h_r\frac{ \mathcal{F}\{S_{pilot}^{\text{affine}}\}}{\sqrt{N}} (m)e^{-\frac{j2\pi m l_r}{N}},~  [m]_{c_1'} = 0. 
\label{equ:y_approach1_pilot}
\end{equation}
Therefore, in this approach, any conventional frequency-domain channel estimation and equalization method can be applied to the combined data, allowing the private data to be extracted directly in the frequency domain, while the common data can be retrieved by converting the equalized signal into the affine domain as shown in the Fig.\ref{fig4}. Alternatively, channel estimation and equalization can be performed in the affine domain, where the common data is directly extracted, and the private data is recovered after transforming the equalized signal back into the frequency domain as depicted in Fig.\ref{fig4}. However, for the 2nd approach, the pilots $Y2_p(m)$ in the received signal is given by 

\begin{equation}
\begin{aligned}
   Y2_{pilot}^{\text{freq}}(m) =\sum_{r=0}^{R-1} h_r&\left(\frac{ \mathcal{F}\{S_{pilot}^{\text{affine}}\}}{\sqrt{N}} (m)+ \frac{ \mathcal{F}\{S_{ce}^{\text{affine}}\}}{\sqrt{N}} (m)\right)
   \\& \times e^{-\frac{j2\pi m l_r}{N}},~  [m]_{c_1'} = 0 .
   \end{aligned}
\label{equ:y_approach2_pilot_freq}
\end{equation}
which shows that the received pilots are not clean and has data contribution to it which will require channel estimation in the affine domain where the extra data and pilots are not overlapping as depicted in Fig. \ref{fig3}. The affine domain received pilots can be given as 
\begin{equation}
\begin{aligned}
   Y2_{pilot}^{\text{affine}}((m) =\sqrt{\phi}\sum_{r=0}^{R-1} h_r P^{\text{affine}} (i-c_1'l_r)
    \\~  [i]_{c_1'} = 0 ~\text{and}~ G<i<N-G,
   \end{aligned}
\label{equ:y_approach2_pilot_affine}
\end{equation}
which allows for conventional AFDM channel estimation \cite{9473655}
Finally, the channel can be equalized either in frequency or in affine domain and the data can be recovered based on the power allocation.

\begin{figure*}[t]
    \centering
        \subfigure[]{\includegraphics[width=0.325\textwidth]{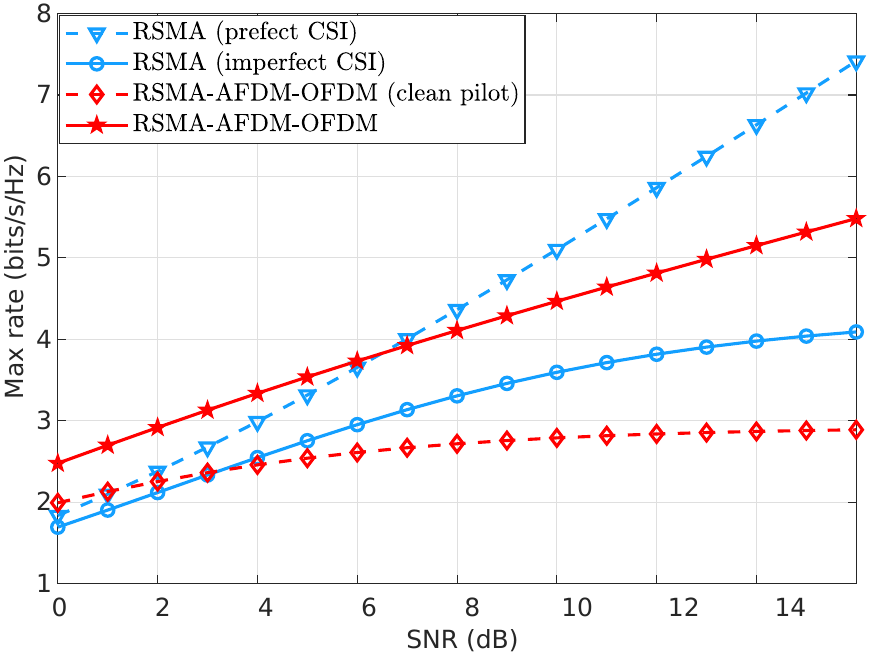}\label{fig5}}
   \subfigure[]{\includegraphics[width=0.325\textwidth,height=4.47cm]{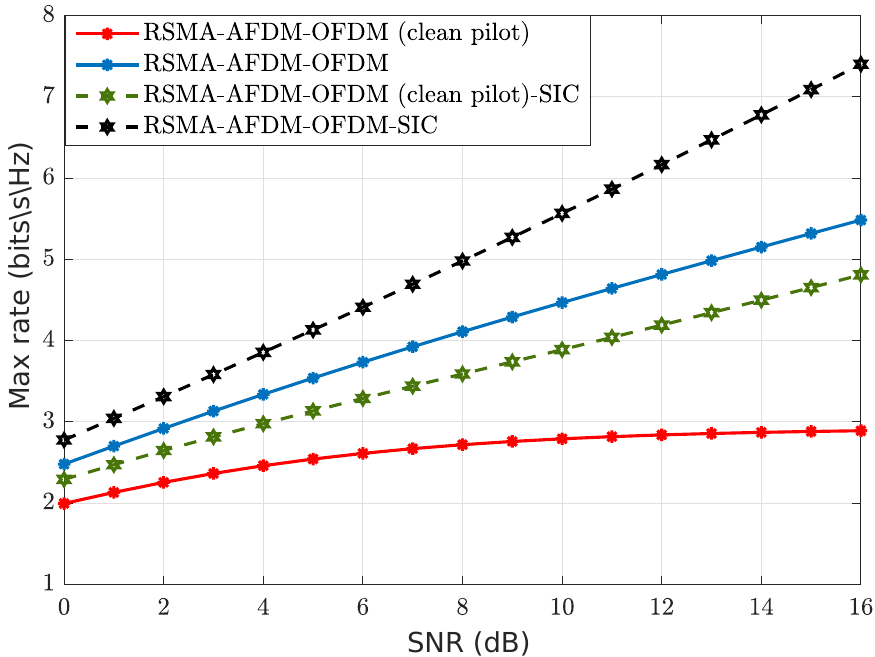}\label{fig6}} 
    \subfigure[]{\includegraphics[width=0.325\textwidth]{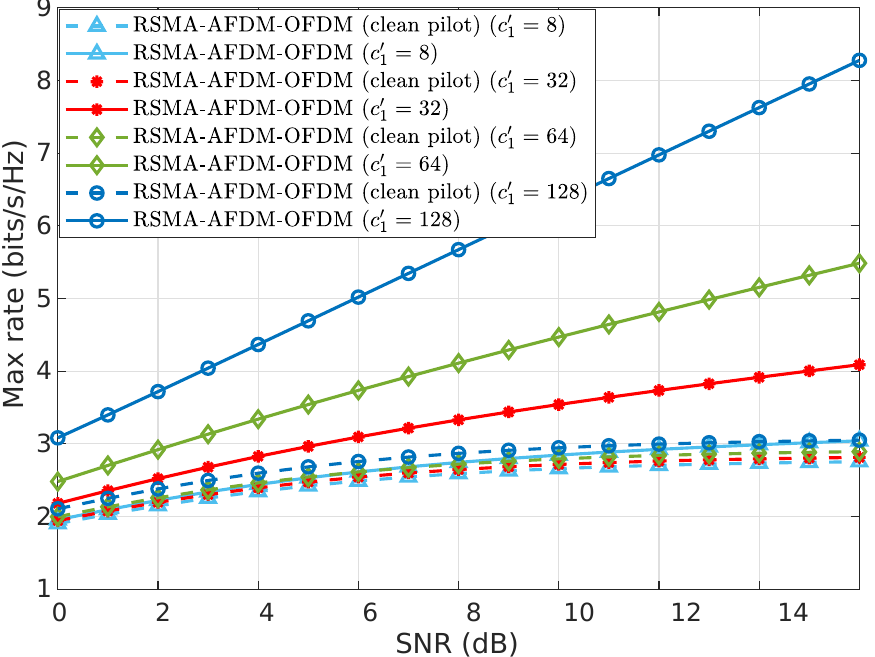}\label{fig7}}

    \caption{(a) SE of the proposed approaches with $c_1' = 64$ vs conventional RSMA, (b) SE of the proposed approach in cases of applying SIC at the receiver with $c_1' = 64$, (c) SE with different values of $c_1'$ .}
    \label{fig:SE}
\end{figure*}
\subsection{Doubly-dispersive channel} 
In a doubly dispersive channel, the Doppler affects the received signal in the frequency domain by introducing a shift leading to pilot-to-data interference. The received signal in approach 1 is given by

\begin{multline}
   Y1_{comb}^{\text{freq}}(m) =\sum_{r=0}^{R-1} h_r S_{\text{comb1}}^{\text{freq}}(m-k_r) \\
   \times e^{-j2\pi \frac{l_r(m-k_r)}{N}},~  [m]_{c_1'} = 0 ,
\label{equ:y_approach1_pilot_dop}
\end{multline}
this shift in the subcarriers will occur also for approach 2. On the other hand, the spreading property of OFDM in the affine domain results in less contribution to guard after the Doppler shifts. Therefore, channel estimation is most effectively conducted in the affine domain for both receiver methods with slight interference from OFDM data to the pilot in both approaches, which can be given for approach 1 as 
\begin{equation}
\begin{aligned}
   Y1_{pilot}^{\text{affine}}((m) =&\sum_{r=0}^{R-1} h_r\big( \sqrt{\phi}P^{\text{affine}} (i-c_1'l_r-k_r)\\+& \mathcal{F}^{-1}\{D_{p1}^{freq}\}(i-c_1'l_r-k_r)\big)e^{-j2\pi \frac{k_r(i-k_r-l_r)}{N}},
   \end{aligned}
\label{equ:y_approach1_pilot_affine}
\end{equation}
where $\mathcal{F}^{-1}\{\cdot\}$  represents the transformation from the frequency domain to the affine domain. Finally, the estimated channel is utilized to equalize the affine domain signal, allowing the common data to be directly extracted, while the private data is recovered by converting the equalized signal back to the frequency domain.

\section{ANALYTICAL AND SIMULATION RESULTS}
This section presents the performance evaluation of the proposed AFDM-OFDM RSMA scheme, comparing it with conventional RSMA and analyzing its effectiveness under various conditions, including the impact of pilot signals, SIC, and waveform spreading factors. The simulation considers SE and BER across different SNRs ranging from 0 dB to 25 dB in steps of 5 dB. The system configuration includes \( N =256\) subcarriers, and QPSK modulation (\( M_{\text{mod}} = 4 \)), with waveform spreading factor for AFDM defined as \( c_1 = \frac{c_1'}{2 M N} \) with \( c_1' \in \mathcal{Z} \). The simulation consists of generating AFDM and OFDM signals, integrating pilot signals for channel estimation, and forming a combined signal that includes AFDM, OFDM, and pilot components. The channel model incorporates two delay taps with Doppler shift enabled, while channel estimation and equalization are applied at the receiver.

\subsection{SE of the proposed approach.}
Figure \ref{fig5} presents the maximum achievable rate (bits/s/Hz) vs. SNR (dB) for different RSMA approaches, highlighting the performance of the proposed AFDM-OFDM RSMA. The RSMA with perfect CSI (blue triangles) achieves the highest rate, reaching approximately 7.5 bits/s/Hz at 16 dB, while RSMA with imperfect CSI (blue circles) suffers from a 46.67\% lower rate, saturating around 4.2 bits/s/Hz at high SNRs. The RSMA-AFDM-OFDM with clean pilot (red diamonds) maintains a nearly constant rate of ~2.9 bits/s/Hz, performing 47.27\% lower than the RSMA-AFDM-OFDM (red stars), which reaches ~5.5 bits/s/Hz at 16 dB, showing an 89.66\% improvement over the clean pilot method. These results confirm that AFDM-OFDM RSMA significantly outperforms RSMA with imperfect CSI, providing a robust SIC-free alternative with higher efficiency for 6G networks.

Figure \ref{fig6} illustrates the maximum achievable rate (bits/s/Hz) vs. SNR (dB) for different RSMA-AFDM-OFDM approaches, comparing the effects of SIC and pilot methods. The RSMA-AFDM-OFDM with clean pilot (red line) achieves the lowest rate, stabilizing around 3 bits/s/Hz at 16 dB, while the RSMA-AFDM-OFDM without SIC (blue line) shows an improvement, reaching approximately 5.5 bits/s/Hz at the same SNR. The RSMA-AFDM-OFDM with clean pilot and SIC (green dashed line) exhibits a 30-40\% gain over the clean pilot method without SIC, showing the impact of SIC on performance. The RSMA-AFDM-OFDM with full SIC (black dashed line) achieves the highest rate, exceeding 7 bits/s/Hz at 16 dB, demonstrating nearly a 50\% increase over its non-SIC counterpart. These results confirm that introducing SIC significantly enhances SE, while the AFDM-OFDM structure without SIC still provides a strong alternative with reduced complexity, making it a viable choice for 6G~networks.

Figure \ref{fig7} presents the SE vs. SNR (dB) for RSMA-AFDM-OFDM under different values of $c_1^{\prime}$, illustrating the impact of waveform spreading factor on performance. The RSMA-AFDM-OFDM (clean pilot) cases (dashed lines) show lower SE compared to their non-clean pilot counterparts, due to the overhead introduced by pilot allocation. As $c_1^{\prime}$ increases from $8$ to $128$, performance improves significantly because of adding more data in the indices $[i]_{c_1^{\prime}}=0$ for approach 2. For instance, at 16 dB SNR, $c_1^{\prime}=128$ (blue circles) achieves nearly 8 bits/s/Hz, whereas $c_1^{\prime}=8$ (triangles) remains below 3 bits/s/Hz, reflecting a $~67$\% increase in SE. This trend confirms that higher $c_1^{\prime}$ values along with utilizing the indices $[i]_{c_1^{\prime}}=0$ enhance spectral efficiency, enabling better resource utilization and signal robustness, making RSMA-AFDM-OFDM an adaptable and scalable multiple access solution for 6G networks.

\subsection{BER of the proposed approach.}
\begin{figure}[t]
    \centering
    \includegraphics[width=1.\linewidth]{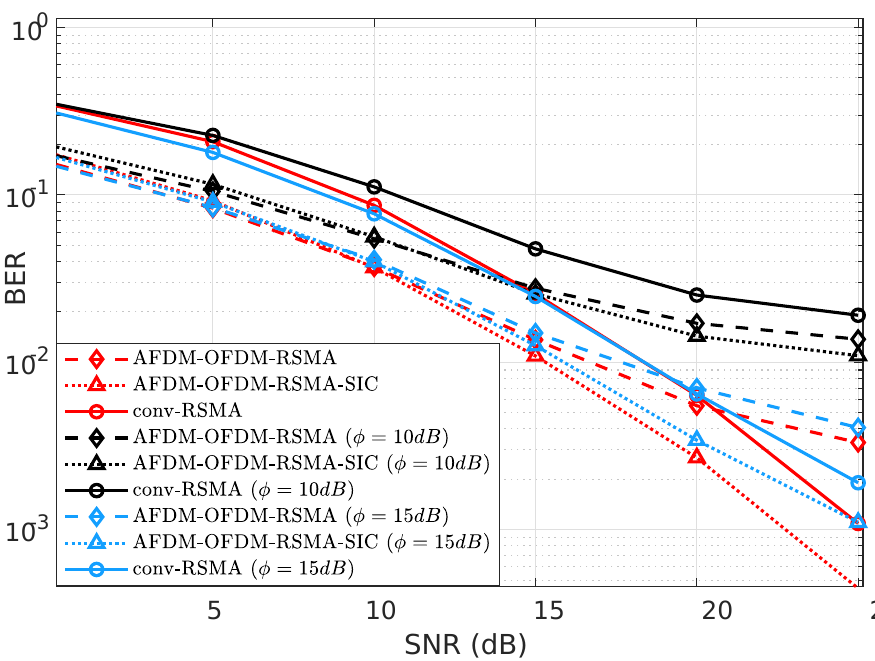}
    \caption{BER of the proposed approach in case of delay~only.}
    \label{fig8}
\end{figure}

Figure \ref{fig8} illustrates the BER vs. SNR (dB) for the proposed AFDM-OFDM-RSMA approach under delay-only conditions, comparing it with conventional RSMA and the effect of SIC. The AFDM-OFDM-RSMA (red diamonds) consistently outperforms conventional RSMA (black circles) across all SNR values. The introduction of SIC (dashed lines) further reduces BER, showing a notable gain, especially at higher SNRs. The effect of power allocation off the pilot ($\phi = 10dB$ vs. $\phi = 15dB$) is also analyzed, with higher pilot power ($\phi = 15dB$) leading to a ~40\% reduction in BER at SNR  $>15 dB$. The best performance is achieved by AFDM-OFDM-RSMA with SIC and ($\phi = 15dB$) (blue triangles), confirming that waveform coexistence and SIC significantly enhance reliability in this delay-only case, making AFDM-OFDM-RSMA a good candidate for 6G networks.

\begin{figure}[t]
    \centering
    \includegraphics[width=1\linewidth]{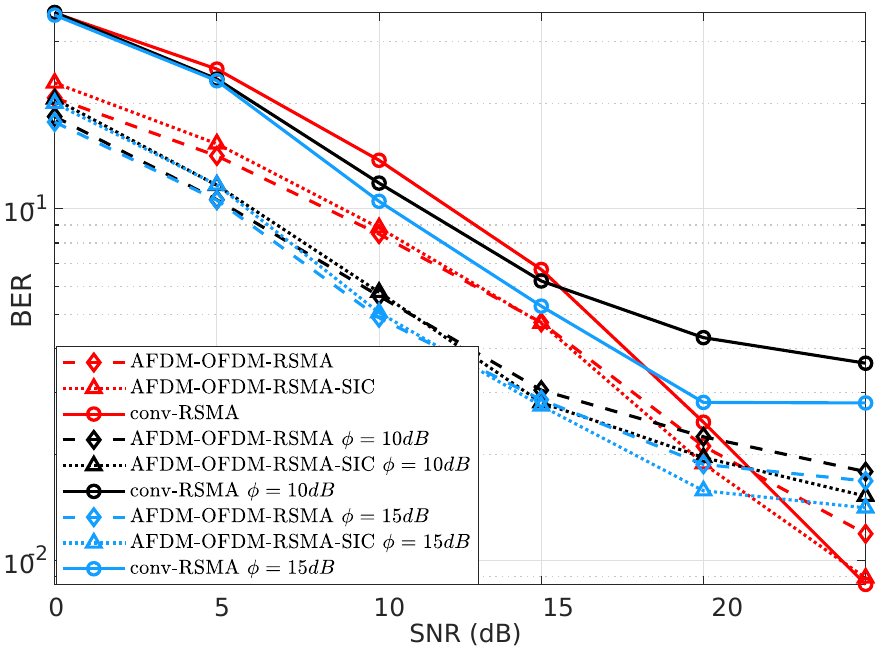}
    \caption{BER of the proposed approach in case of delay and~Doppler.}
    \label{fig9}
\end{figure}

Figure \ref{fig9} illustrates the BER vs. SNR (dB) for the proposed AFDM-OFDM-RSMA approach under delay and Doppler conditions, comparing it with conventional RSMA and evaluating the impact of SIC. The AFDM-OFDM-RSMA (red diamonds) consistently outperforms conventional RSMA (black circles), demonstrating better resilience against Doppler-induced distortions. The introduction of SIC (dashed lines) further improves BER performance, particularly at SNR > 10 dB, confirming its effectiveness in mitigating interference. The impact of pilot power ($\phi = 10dB$ vs. $\phi = 15dB$)  is also analyzed, showing that higher pilot power ($\phi = 15dB$) reduces BER by approximately 50\% at SNR = 20 dB. The best performance is observed for AFDM-OFDM-RSMA with SIC and ($\phi = 15dB$) (blue triangles) also in this case.
Comparing the delay-only case Fig. \ref{fig9} and the delay-Doppler case Fig. \ref{fig9}, we observe that Doppler effects degrade BER performance, especially at lower SNR values.

In Fig. \ref{fig9}, the BER for all schemes is consistently higher compared to Fig. \ref{fig8}, indicating the additional challenge posed by Doppler-induced frequency shifts. However, AFDM-OFDM-RSMA with SIC and higher pilot power ($\phi = 15dB$) still achieves the best BER performance in both cases, showing its robustness in dynamic environments. Notably, the performance gap between conventional RSMA and AFDM-OFDM-RSMA widens in the delay-Doppler scenario, further highlighting the superiority of AFDM-OFDM in handling time-frequency dispersion effects.

\subsection{Complexity Analysis}
The computational complexity of the proposed AFDM-OFDM-based RSMA receiver is compared with the conventional RSMA receiver, focusing solely on the operations that differ between the two while excluding common steps. The primary distinction lies in the decoding strategy employed: the conventional RSMA receiver relies on SIC, which incurs significant computational overhead. SIC involves iterative decoding and interference subtraction, which depend on the number of users and the number of antennas. The total complexity of RSMA $C_{RSMA}$ considering an Maximum Likelihood (ML) detector, can be written as \cite{10508205}
\begin{equation}
C_{RSMA} =  N \sum_{k=1}^{2} 
\underbrace{(4C M + 2 M^C)}_{\text{ML detection}} 
 (2-k+1) 
+ \underbrace{O(M^2)}_{\text{Subtraction}} 
\end{equation}

where $M$ and $C$ are the modulation order and the number of channels between the BS and the $k$-th user, respectively. In contrast, the proposed AFDM-OFDM-based RSMA receiver eliminates the need for SIC by employing a domain-based separation strategy. In this scheme, the common and private data streams are separated into distinct domains: the affine domain for AFDM and the frequency domain for OFDM. This approach replaces the iterative decoding of SIC with DAFT and its inverse (IDAFT), each of which has a complexity $O(2N+N \log N)$, where $N$ is the number of subcarriers. While the introduction of DAFT and IDAFT adds computational overhead, this overhead remains significantly lower than the complexity of SIC, especially for scenarios with a large number of users.

%\newpage
\section{CONCLUSION AND FUTURE WORK}
In this paper, we presented a novel waveform coexistence-driven RSMA framework that leverages the integration of AFDM and OFDM to enhance spectral efficiency, robustness, and reliability in 6G networks. The proposed AFDM-OFDM RSMA eliminates the need for SIC by utilizing AFDM for common data at higher power in the affine domain and OFDM for private data at lower power in the frequency domain, enabling efficient interference management. Simulation results confirm that the proposed system achieves enhanced spectral efficiency in practical scenarios, even with imperfections in channel estimation, making it a promising candidate for NGMA solutions.
Future work includes extending the proposed AFDM-OFDM RSMA framework to multi-antenna (MU-MIMO) scenarios with imperfect CSI, further optimizing power allocation and waveform design to enhance system adaptability. Finally, the integration of RSMA with alternative waveforms will be explored to develop a comprehensive interference-resilient multiple access strategy.

\end{document}